\documentclass[review]{elsarticle}

\usepackage{hyperref}
\journal{Physica D}

\usepackage{amsmath,amsfonts,amssymb}
\usepackage{tabularx}
\usepackage{xcolor}

\newcommand{\zmax}{z_\mathrm{max}}
\newcommand{\tx}{\Tilde{x}}
\newcommand{\ty}{\Tilde{y}}
\newcommand{\tr}{\Tilde{r}}
\newcommand{\tp}{\Tilde{p}}
\newcommand{\td}{\Tilde{d}}
\newcommand{\Tt}{\Tilde{t}}
\newcommand{\LE}{\!\leq\!}
\newcommand{\EQ}{\!=\!}
\newcommand{\IN}{\!\in\!}
\newcommand{\+}{\! + \!}
\renewcommand{\L}{{\mathcal{L}}}

\begin{document}

% = = = = = = = = = = = = = = = = = = = = = = = = = = = = = = = 

\begin{frontmatter}

% ONLY TITLE
\title{The role of bilinguals in the {B}ayesian naming game}

% TITLE + NOTE
%\title{The Role of bilinguals in the Bayesian naming game\tnoteref{mytitlenote}}

\author{Gionni Marchetti\corref{mycorrespondingauthor}}
\ead{gionni.marchetti@kbfi.ee}
\cortext[mycorrespondingauthor]{Corresponding author}
\author{Marco Patriarca}
\ead{marco.patriarca@kbfi.ee}
\author{Els Heinsalu}
\ead{els.heinsalu@kbfi.ee}
\address{National Institute of Chemical Physics and Biophysics, R\"avala 10, 10143 Tallinn, Estonia}

% NOTE \fnref{...}
% Ex.:  \author{Gionni Marchetti}\fnref{myfootnote}}
%       \fntext[myfootnote]{Since 1880.}


\begin{abstract}
We study the recently introduced Bayesian naming game model, in which the one-shot learning of the minimal naming game is replaced by a more realistic learning process defined according to Bayesian inference.
The results are compared with those obtained from the minimal naming game model.
We focus on the dynamics of the bilingual population, providing analytical estimates of the upper bound for the number of bilinguals in both models based on the mean-field equations, and validate them through numerical simulations of the multi-agent models.
We show that in the Bayesian model the maximum number of bilinguals is always lower with respect to the minimal naming game and that the two models are characterized by qualitatively different time evolutions.
%We discuss these results from a geometrical point of view, by considering the trajectory in the phase plane and tracing back the differences of the BNG to the additional random fluctuations induced by the underlying learning process.
\end{abstract}


\end{frontmatter}

% = = = = = = = = = = = = = = = = = = = = = = = = = = = = = = = 

%\linenumbers

% = = = = = = = = = = = = = = = = = = = = = = = = = = = = = = = 

\section{Introduction}

Understanding how consensus spontaneously emerges in various contexts (language, money, dress codes, etc.) \cite{Erlich2005, Nyborg2016, Baronchelli-2018a}  is of theoretical and practical importance and has been studied within different disciplines including complex systems theory.
Among other approaches, the ``naming game'' (NG) is a simple yet insightful model that was introduced by Baronchelli et al.~\cite{Baronchelli-2006c, chen-2019a}, inspired by Wittengstein language games \cite{Wittgenstein-1953} and Steels' works~\cite{Steels-1995a,Steels-1997b}. 
The NG model and its variants~\cite{Baronchelli-2006a, Baronchelli-2007a, Baronchelli-2007b} have been studied in different complex network topologies, 
inter-agent communication protocols, 
learning rules, 
and heterogeneous ensembles of agents~\cite{Marchetti-2020b,Patriarca-2020a}.

In the the minimal NG (MNG), introduced in Ref.~\cite{Baronchelli-2006a} (see also Ref.~\cite{Baronchelli-2016a}), the learning process is deterministic and instantaneous, since it can take place during a single communication.
In order to make the MNG more realistic, in Refs.~\cite{Marchetti-2020a, Marchetti-2020b} we introduced the Bayesian naming game (BNG). 
The model is based on two assumptions suggested by experimental observations.
First, learning words requires multiple cognitive efforts over an extended period of time (iterate learning)~\cite{Tenenbaum-1999, Griffiths-2007}. 
Second, the word-learning process can be seen as an approximate form of the Bayesian inference~\cite{Tenenbaum-1999, Tenenbaum-1999b, Tenenbaum-2000b, Tenenbaum2001, Xu-2007a, Perfors-2011a, Murphy-2012a}.

In the BNG, the agents behave according to the Bayesian learning framework developed by Tenenbaum  and co-workers~\cite{Tenenbaum-1999,Tenenbaum-2011a,Perfors-2011a,Murphy-2012a,Lake2015}.
This provides a richer dynamics with respect to the one observed in the MNG model, despite both  models exhibit a disorder-order transition where consensus emerges spontaneously.
The BNG model provides new tools for investigating the effects of human cognitive biases on social consensus~\cite{Tversky1974,  Hahn2014, Sanborn2016, Ngam2016, madsen2018}: through the Bayes' theorem \cite{Jeffreys1961}, the BNG model takes into account the learner's previous experiences and background knowledge.  
The relevance of the cognitive dimension is apparent, for example, in Refs.~\cite{Baronchelli2010e, Loreto2012, Baronchelli-2015}, where modeling the cognitive (or perceptual) bias in a category-game model (a generalization of the NG model) of colors was the key to obtain a good agreement with the color hierarchy reported in the World Color Survey.

%%% ABOUT THIS PAPER

In this paper, we address the dynamics of the MNG and the BNG models restricted to two names A and B.
These models can be understood as  three-state models~\cite{Patriarca-2012a,Patriarca-2020a}, since there can be monolingual agents that know only name  A or B and bilingual agents that know both names A and B.
Rather than considering the population fractions of monolinguals, as often done, in this paper we shall focus on the dynamics of the bilingual agents for two main reasons:
In fact,  the bilingual agents play a key-role in the dynamics, as they constitute the required intermediate state for A$\leftrightarrow$B transitions necessary for reaching the consensus~\cite{Castello2006a, Marchetti-2020a};
the study of their dynamics highlights some crucial differences between the two multi-agent models considered, which is expected to have relevant consequences in more general models describing the competition of multiple words, as discussed below.

The paper is structured as follows.
In Sec.~\ref{sec:NGBNG} we present the MNG and BNG models restricted to two names A and B.
In Sec.~\ref{sec:consideration} we show that, through a suitable coordinate transformation, the study of the fraction of bilinguals can be reformulated in terms of a single ordinary differential equation, which resembles a Riccati equation.
Using this equation, we obtain an upper bound for the maximum fraction of bilinguals emerging during the semiotic dynamics.
In Sec.~\ref{sec:results}, through numerical simulations, we carry out a detailed comparison between the MNG and the BNG models.
%due to the replacement of the one-shot name learning by a real-setting name learning based on Bayesian inference
First, we validate our estimate of the maximum number of bilinguals.
Furthermore, we show that the time evolutions of the MNG and BNG are characterized by some relevant differences that can be traced back to the different name learning processes.
Finally, in Sec.~\ref{sec:conclusion} we discuss the results obtained and their importance for modeling the word learning process and the emergence of consensus.

 % --------------------------------------------------------------
\begin{figure}[ht!]
    \resizebox{0.75\textwidth}{!}{%\includegraphics{phasePortraitNew.pdf}
        \includegraphics{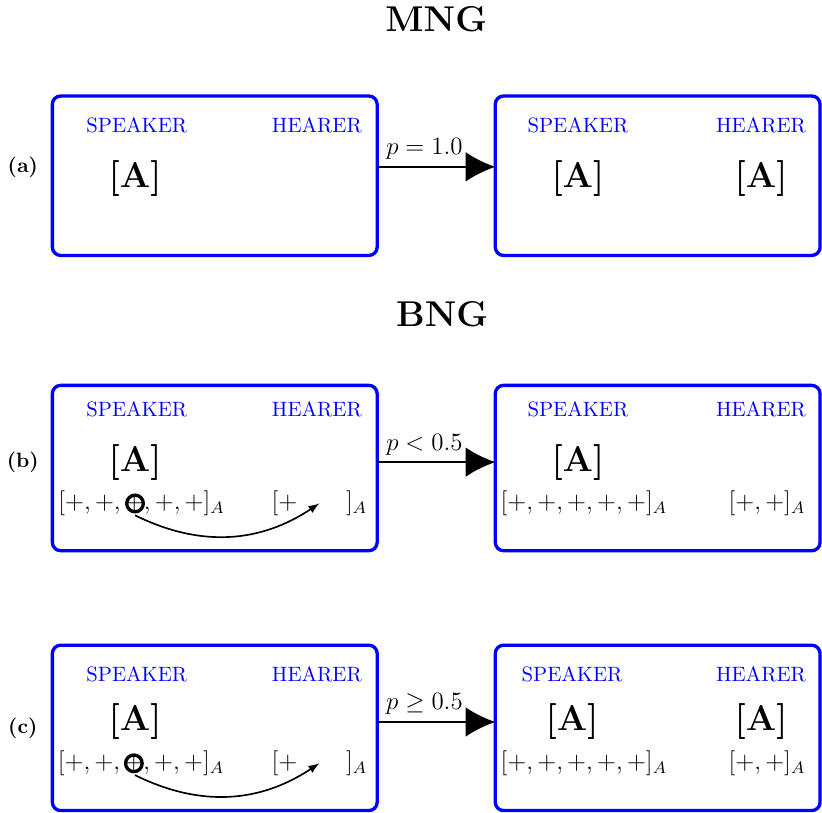}
    }
    \caption{Learning processes in the MNG and BNG models (for the sake of clarity only the uttered word A is represented).
        Panel (a): MNG model. 
        The speaker $i$ utters the name A, which is not present in the hearer's list $\L_j$. 
        Next, the hearer $j$ learns A with probability $p=1$ (one-shot learning) and hence adds A to the list $\L_j$.
        Panels (b) and (c): BNG model.  
        The speaker $i$ conveys name A together with the corresponding example ``+'', randomly chosen in the inventory, to the hearer $j$. 
        The hearer $j$  tries to learn it by computing the generalization probability $g$ by means of Eq.~\eqref{eq:generalization}. 
        If $p \le p^* = 0.5$ the hearer does not learn the new word A and only records the positive example ``+'' in the inventory (panel (b)). 
        Instead, if  $p \ge p^* = 0.5$ the learning process is successful and the hearer $j$, besides adding the corresponding example ``+'' to the inventory $[+++\dots]_A$, adds also the name A to the to the name list $\L_j$ (panel (c)).}
    \label{fig:rules}
\end{figure}
% ---------------------------------------------------------------- 

% = = = = = = = = = = = = = = = = = = = = = = = = = = = = = = = 

\section{The minimal and Bayesian naming game models} 
\label{sec:NGBNG}

The NG was introduced for modeling the spontaneous emergence of consensus through mutual interactions in a population of agents, to understand if and how a set of agents reaches an agreement about which name to use for referring to a certain object.
Other semiotic dynamics models describe agents naming a set of different objects at the same time~\cite{Hurford-1989a,Nowak-1999a}, which is a more realistic situation.
However, considering a single object as done in the NG model does not exclude that agents have already named other objects; it only implies the simplifying assumption that the convergence to consensus on how to name a certain object is not influenced by other objects.
A scenario described by the NG is that of a group of agents that have already agreed on how to name a set of objects, but need to name a new object that they see for the first time.

% --------------------------------------------------------------

\subsection{The minimal naming game model} 
\label{sec:NG}

In the MNG model, each agent $i$ ($i=1,\dots,N$) is provided with a list $\L_i$ of known synonyms.
In the present paper we study the case of a single object associated to two synonyms A and B.
Therefore the list of an agent can be [A], [B], or [A,B].

During the semiotic dynamics of the MNG,  at each time-step two agents $i, j$, the speaker and the hearer respectively, are randomly chosen. 
The speaker $i$ selects a name, A or B, present in the name list  $\L_i$, randomly choosing one of the two if both are present.
If the uttered name is present in the hearer's name list  $\L_j$, both agents update their lists keeping the uttered name  only. 
When this agreement process happens, the interaction is considered successful. 
Otherwise, when the word conveyed by the speaker is not present in the hearer's name list, the latter adds it to $\L_j$, see Fig.~\ref{fig:rules}-(a).
%(note that this is considered a failure~\cite{Baronchelli-2006c}). 
Here we refer to this process as one-shot learning process, since in this case the hearer learns the name at the first attempt with probability $p=1$ \footnote{Note that in cognitive science one-shot learning usually refers to generalizing a concept after a few attempts. This is indeed the case of the BNG model.}. 
The time $t$ is measured as the number of time steps elapsed, i.e. the number of interactions between pairs of agents.

 % --------------------------------------------------------------
%\begin{figure}[ht!]
%    \resizebox{0.75\textwidth}{!}{%\includegraphics{phasePortraitNew.pdf}
%        \includegraphics{positiveExample.png}
%    }
%    \caption{The name learning in a real-life setting. In the cartoon the child and her mother are the hearer and speaker, respectively, according to the BNG model. The above  event illustrates a positive example by which the child learns the word  ‘cat’. This training event experienced by the hearer is customarily denoted by the symbol ``+''. All the training events similar to the one illustrate above, are  added by the hearer to her  inventory  $[+++\dots]$ corresponding to the name  ‘cat’. This cartoon was generated from the website  https://www.makebeliefscomix.com.}
%    \label{fig:positiveExample}
%\end{figure}
% ---------------------------------------------------------------- 

% ------------------------------------------------------------

\subsection{The Bayesian naming game model} 
\label{sec:BNG}

Before proceeding with the BNG model in its technical details, let us discuss its origin and the principal differences with respect to the MNG.

In the late 90's Tenenbaum developed a computational theory of concept learning based on Bayesian inference~\cite{Tenenbaum-1999}, which he later applied to the problem of learning novel words from examples~\cite{Xu-2007a, Tenenbaum-2000b}. 
Tenenbaum's goal was to explain within a Bayesian learning framework how people can generalize meaningfully from  a few positive examples of a new word. According to this cognitive computational approach the human-like learning of a new word also implies the learning of its meaning (or the concept $\mathcal{C}$ associated to it).

Recently  we incorporated this human-like word learning into the MNG model ~\cite{Marchetti-2020a, Marchetti-2020b}. 
To this end, we replaced the typical one-shot learning of the MNG model with Tenenbaum's learning framework. 
The aim of the proposed BNG is to simulate real-life situations, similar to those of a child who is trying to learn the meaning of the word “cat” after experiencing a small number of training events, e.g., when the mother says “This is a cat” while pointing to a cat. 
After a few similar training processes we would expect that the child will associate the meaning of “cat” to all cats independently of their breed (Abyssinian, Siamese, etc.) at the same time excluding horses, dogs, etc.

Note that the aforementioned training events experienced by the hearer are called the positive examples corresponding to the learning process of an unknown word  (“cat” in our example). 
The examples are denoted  by the symbol ``+''~\cite{Tenenbaum-1999}. 
In both models the object to be named (e.g. an animal, a toy, etc.) is incidental and is usually assumed to belong to the common environment where the agents perform their pairwise interactions. 
In this regard,  we shall assume that the BNG model takes into account the name learning of objects whose meaning (or equivalently concept $\mathcal{C}$) is not too abstract~\footnote{For instance, we exclude mathematical concepts, e.g. the concept of irrational number.}.

\begin{figure}[ht!]
    \resizebox{0.5\textwidth}{!}{%\includegraphics{phasePortraitNew.pdf}
        \includegraphics{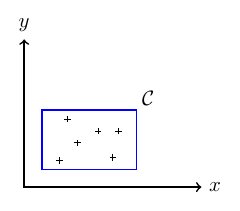}
    }
    \caption{Example of concept $\mathcal{C}$ represented as a rectangle in $\mathbb{R}^{2}$, together with six corresponding positive examples ``+''.
    See also Ref.~\cite{Tenenbaum-1999b}.}
    \label{fig:rectangle}
\end{figure}

Next, the Bayesian theory developed by Tenenbaum requires that the positive examples must be uniformly sampled at random from the concept $\mathcal{C}$ to be learned by the agents. 
To this end,  we shall assume  that a given concept $\mathcal{C}$ can be represented as a axis-parallel rectangle
%, i.e., a rectangle concept, 
in the Cartesian plane $\mathbb{R}^{2}$,  see Fig.~\ref{fig:rectangle}. 
According to this choice it is easy to uniformly generate the positive examples at random inside a given rectangle in  $\mathbb{R}^{2}$ -- see below.
%(see Eq.~\ref{eq:generalization} and corresponding computational details in the next paragraphs). 
Despite this rather abstract approach for modelling a concept, it is worth noting that some concepts may have the above geometric representation~\cite{Tenenbaum-1999b}. 
For example, considering the concept of health of an individual related to insulin and cholesterol levels $x$ and $y$, the two specific intervals of real numbers, in which $x$ and $y$ have to be, determine a rectangle in the $x$-$y$ plane, as depicted in Fig.~\ref{fig:rectangle} (see also Ref.~\cite{Tenenbaum-1999b}).

% --------------------------------------------------------------
% ---------------------------------------------------------------- 

Incorporating the Bayesian learning framework into the minimal naming semiotic model comes at the cost of introducing additional uncertainty into the pairwise interactions between the agents. 
This uncertainty arises from the probabilistic nature of the Bayesian inference. 
However, it is worth noting that in everyday life reasoning is characterized by some uncertainty and the BNG model represents a first step towards semiotic dynamics models that describe human-like learning~\footnote{If we ignore  the above mentioned cognitive framework, we may think of the uncertainty in the BNG model as simply arising from some kind of noise  present during the pairwise interactions between the agents.}.

Let us now discuss the mathematical details of the BNG model.
Within the BNG model, a Bayesian agent learns a name generalizing the concept $\mathcal{C}$  associated to it. 
Here a given concept $\mathcal{C}$ is represented by a tuple $\left( x, y,  \sigma_1,  \sigma_2 \right)$ defining an  axis-parallel rectangle in $\mathbb{R}^{2}$, where $x, y$ are the lower-left corner's  coordinates and  $\sigma_1,  \sigma_2$ the rectangle's sizes along the $x$ and $y$-axis, respectively. 
All the axis-parallel rectangles that can be drawn in the plane constitute the hypothesis space $\mathcal{H}$. 
The Bayesian agent can generalize the concept  from a new positive example $\xi$, which is a point in the $x$-$y$ plane, if the generalization function $g$ obtained by Bayes' theorem is $g > p^*$, where $p^*$ is a given threshold probability for generalizing $\mathcal{C}$, and $g$ is given by  (see Ref.~\cite{Tenenbaum-1999} for details)
%Ref.~\cite{Jeffreys1961, Stone2013, Tenenbaum-1999},
%
\begin{equation}\label{eq:generalization}
    g \left( \xi \in \mathcal{C} | X \right) 
    \approx 
    \frac{\exp \left \{ -  \left(   \frac{\td_1 }{\sigma_1} +     \frac{\td_2 }{\sigma_2 }      \right)   \right \} }
              {\left[ \left( 1 + \frac{\td_1 }{r_1} \right) \left( 1 + \frac{\td_2 }{r_2} \right) \right]^{n - \alpha} }  \, .
\end{equation}
Here, $X$ denotes the set of the previous examples recorded by the agent and $n$ is the number of examples contained in it.
The quantity $r_i$ is the maximum distance along $x$-axis ($i = 1$) or $y$-axis  ($i =  2$) between the examples in $X$, while $\td_i$ ($i=1,2$) is zero if the $i$-th coordinates of the new example $\xi$ falls inside the relative interval of values determined by $X$ or otherwise it equals the distance from the nearest example belonging to $X$.
The symbol $\alpha$ stands for the chosen  prior density, as it seems reasonable to assume that the agents have some qualitative  knowledge of the possible size of the concept $\mathcal{C}$; in this paper most of computations will be performed using the Erlang prior, corresponding to $\alpha = 2 $, when the agents might expect that concepts with size much smaller or larger than those of $\mathcal{C}$ are extremely rare~\cite{Tenenbaum-1999}. 
Instead, in the case of the exponential prior (or the maximum entropy density), corresponding to $\alpha = 1$, agents know the expected size of $\mathcal{C}$.
In Sec.~\ref{sec:robust} we shall show that when the exponential prior is adopted for the problem at hand, no major differences are found.

From a computational point of view, the Bayesian cognitive model can be embedded in the naming game in the following way. 
First, beside the name list $\L_i$, each agent  is equipped with two additional inventories, $[+++\dots]_A$ and $[+++\dots]_B$, where the ``+'' symbols represent all the positive examples corresponding to the names A and B, respectively, recorded by each agent. 
During an encounter, speaker $i$ utters a name, for example A as in Fig.~\ref{fig:rules}-(b)-(c), conveying at the same time also a positive example ``+'' to hearer $j$, who records the example in the corresponding inventory.
Hearer $j$ computes the probability $p = g(\xi \in \mathcal{C} | X)$, where $g$ is defined in Eq.~\eqref{eq:generalization} and $\xi$ represents the new unseen example delivered by the speaker $i$ along with the uttered name. 
If the computed probability $p > p^*$, where  $p^*$ is the threshold probability for generalizing $\mathcal{C}$, the generalization is successful and hence the hearer adds the corresponding name to the name list $\L_j$, see Fig.~\ref{fig:rules}-(c);
otherwise nothing happens, apart that the new example $\xi$ is recorded in the inventory of the hearer --- we refer to the latter event as reinforcement process, Fig.~\ref{fig:rules}-(b).
It is customary to assume a threshold $p^* = 0.5$~\cite{Tenenbaum-1999}. 
%{The schemes of the different learning processes in the MNG and BNG models are compared in Fig.~\ref{fig:rules}.}

In this work the positive examples will be points generated at random within the axis-parallel rectangle $\left( 0, 0,  3,  1 \right)$ according to the the Bayesian strong sampling assumption \cite{Tenenbaum-1999, Marchetti-2020a, Marchetti-2020b}. 
However, this particular choice does not affect the numerical results. 

Our multi-agent simulations start from initial conditions in which there are only monolingual agents. 
Half of them have name A and the other half name B in their name lists and the corresponding inventories $[+++\dots]_A$ and $[+++\dots]_B$ contain $n_{ex}=4$ examples.
When a hearer, e.g. an A-monolingual, has an inventory with at least $n^{\ast}_{ex}=5$ examples for the unknown word B, the hearer can start generalizing the concept $\mathcal{C}$ in association with B \footnote{In the original BNG model \cite{Marchetti-2020a, Marchetti-2020b} an initial bias was assumed in order to make the synonyms A and B distinguishable \cite{Edmonds2002}, by assigning different thresholds $n^{\ast}_{ex, A}=5$ and $n^{\ast}_{ex, B}=6$. In this paper equal thresholds are assumed to ensure a fair comparison between BNG and MNG models.}.

% = = = = = = = = = = = = = = = = = = = = = = = = = = = = = = = 

\section{Upper bound for bilinguals: Mean-field approximation} 
\label{sec:consideration}

\subsection{Mean-field dynamics}
\label{sec:meanfield}

The mean-field dynamics of the MNG and BNG models described in the previous section can be formulated in a unified way. 
Denoting the population fractions of agents that know only name A or only name B at time $t$ by $x(t)$ and $y(t)$,  respectively, and the fraction of bilinguals by $z(t) = 1 - x(t) - y(t)$, the mean-field equations read~\cite{Marchetti-2020a}
\begin{align}
  \dot{x} = - p_B x y + \left(1- x - y \right) ^{2} + \frac{3 - p_B}{2} x \left(1- x - y \right)  \, , \label{eq:autonomous1a}\\  %\nonumber
  \dot{y} = - p_A x y +  \left(1- x - y \right)^{2} +  \frac{3 - p_A}{2} y \left(1- x - y \right) \label{eq:autonomous1b} \, ,
\end{align}
where $p_A(t), p_B(t) \!\in\! [0,1]$ represent the probabilities at time $t$ that a monolingual of B or A generalizes the concept in association with name A or B, respectively, see Ref.~\cite{Marchetti-2020a} for details.
The quantities $p_A(t)$ and $p_B(t)$ are obtained in the mean-field limit as ensemble averages over the monolinguals of B and A, respectively, of the generalization function defined in Eq.~\eqref{eq:generalization}~\cite{Marchetti-2020a}.
One can expect that $p_A(t), p_B(t)$ depend nonlinearly on the history of the system, so that they have a complex (unknown) functional dependence on the population fractions.
What we know is that in the initial state $p_A(0) = p_B(0) = 0$, since the agents have no examples of the unknown word (see previous section);
as more and more agents learn the initially unknown word, $p_A(t)$ and $p_B(t)$ grow monotonously, until the value $1$ is reached.

% --------------------------------------------------------------
\begin{figure}[ht!]
    \resizebox{0.95\textwidth}{!}{%\includegraphics{phasePortraitNew.pdf}
        \includegraphics{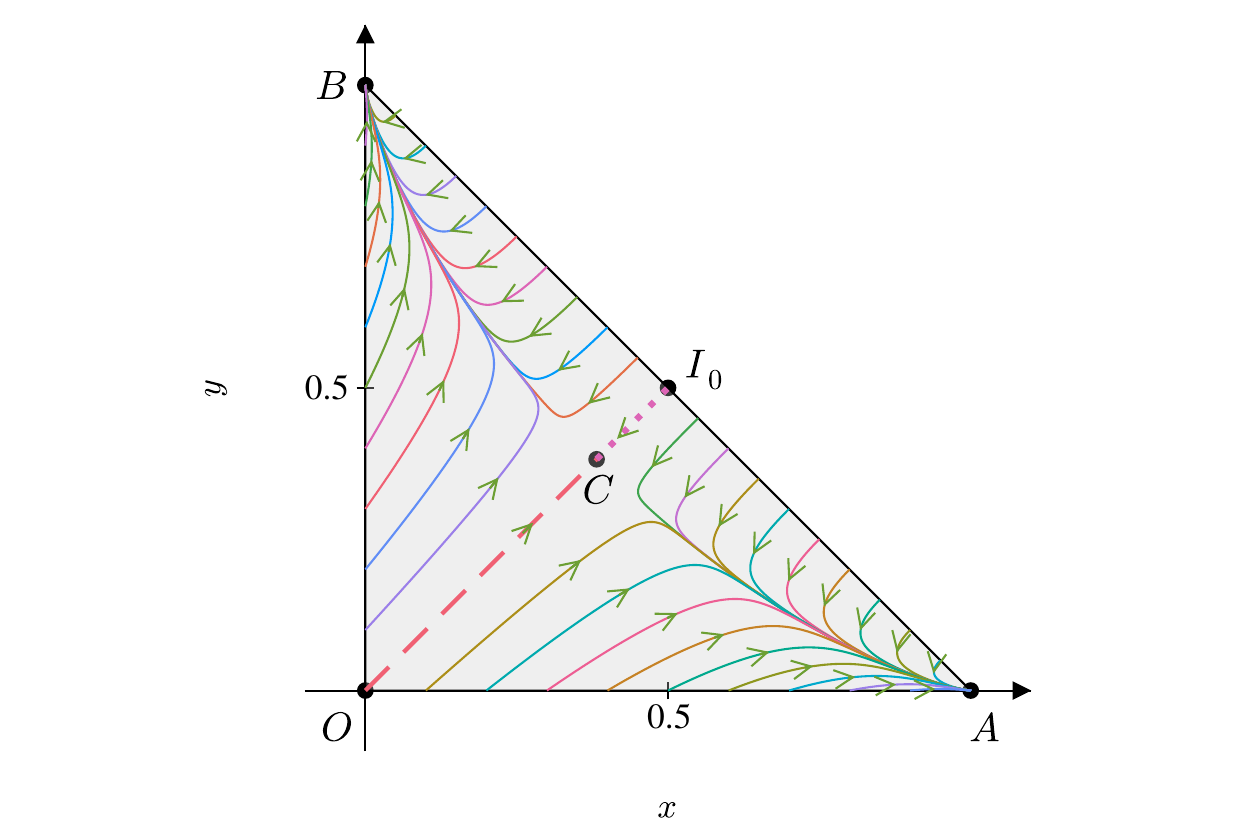}
    }
    \caption{
    Phase portrait of the MNG model with the streamlines induced by the velocity field provided by Eqs.~\eqref{eq:autonomous1a}, \eqref{eq:autonomous1b} with $p_A = p_B = 1$;
    trajectories are allowed to roam within the region of the phase space defined by the constraints 
    $x \ge 0$, $y \ge 0$ and $z = 1 - x - y \ge 0$ (shadowed area).
    The points $A=(1,0)$ and $B=(0,1)$ are asymptotically stable equilibrium points, 
    while $C= \left((3-\sqrt{5})/2, (3-\sqrt{5})/2\right)$ is an unstable saddle point. 
    The dotted and dashed lines represent the trajectories starting from $I_0 = (0.5,0.5)$ and from $O = (0,0)$, respectively; 
    in these cases the representative point remains on the $x=y$ line and asymptotically reaches point $C$.
    This phase portrait shares various features with that of the BNG model, see text for details.
}
    \label{fig:phasePortrait}
\end{figure}
% ---------------------------------------------------------------- 

The MNG model is recovered for $g = 1$; in this case $p_A(t) = p_B(t) = 1$ at any time $t$ and Eqs.~\eqref{eq:autonomous1a},~\eqref{eq:autonomous1b} become an autonomous nonlinear system of first-order differential equations. 
The MNG phase portrait is shown in Fig.~\ref{fig:phasePortrait}.
The region of the phase space, where the trajectories are allowed to roam, is defined by the constraints $x,y \ge 0$ and $x + y \le 1$.
There are two stable equilibrium points $A=(1,0)$, $B=(0,1)$ and an unstable equilibrium point $C=((3-\sqrt{5})/2,(3-\sqrt{5})/2)$, as can be shown by linear stability analysis~\cite{Baronchelli-2006a}.
The trajectories in Fig.~\ref{fig:phasePortrait} converge either to $A$ or $B$, apart from those starting from any point lying on the line $x=y$, which converge toward point $C$, including the trajectories starting from the origin $O$ (where only bilinguals are present, $z(0) = 1)$ and from the symmetrical initial condition $I_0 = (0.5,0.5)$  (where there are no bilinguals).
%In fact, point $C$ is a saddle point in the $x$-$y$ plane but represents a stable equilibrium point if the motion is constrained along the $x=y$ line, which happens when $x(0)=y(0)$, as one can check from Eqs.~\eqref{eq:autonomous1a},~\eqref{eq:autonomous1b}.

It is easy to see that in the BNG model $A$ and $B$ are still asymptotically stable equilibrium points.
Since the MNG represents a particular limit of the BNG (for $p_A = p_B = 1$), it is natural to expect that also in the BNG model there should be an additional equilibrium point, analogous to point $C$ of the MNG.
In order to compare directly the BNG with the MNG, it is useful to consider the learning probabilities as effective time-dependent parameters $p_A(t)$ and $p_B(t)$, rather than complex functionals of $x(t)$ and $y(t)$.
Because in the symmetrical initial state $I_0$ one has $x(0) = y(0) = 0.5$ and $p_A(0) = p_B(0) = 0$, since agents have no examples of the unknown word, the probabilities $p_A(t)$ and $p_B(t)$ will change with time exactly in the same way, i.e. $p_A(t) = p_B(t)$.
Thus, in the mean-field limit $x(t) = y(t)$ and one can set $p_A(t) = p_B(t) \equiv p(t)$ in Eqs.~\eqref{eq:autonomous1a},~\eqref{eq:autonomous1b}. 
It can be shown that there exists an unstable equilibrium point $C'(t) = (c_0(t),c_0(t))$, 
where $c_0(t) = \frac{1}{4}\left[p(t) + 5 - \sqrt{p(t)^2 + 10 p(t) + 9~}\right]$.
As for the time-dependence of point $C'(t)$, initially $C'$ coincides with the symmetrical initial condition $I_0$, i.e. $C'(0) = I_0$; then $C'(t)$ moves toward the unstable equilibrium point $C$ of the MNG; eventually $C'(t\!\to\!\infty) \to C$, since for $t\!\to\!\infty$ more and more agents have learned the initially unknown word, which leads to $p_A(t), p_B(t) \to 1$ and therefore to the equivalence between the MNG and BNG dynamics.

A convergence toward an unstable equilibrium point, like the one discussed for the MNG and BNG, is found also in the mean-field solutions of other two- and three-state models of competition between equivalent languages, when starting from symmetrical initial conditions~\cite{Stauffer-2007a,Vazquez-2010a}. 
In a many-agent model this cannot happen due to the random fluctuations and the corresponding solutions are very different, see Sec.~\ref{sec:results}.

%-------------------------------------------------
\begin{figure}[ht!]
    \resizebox{0.75\textwidth}{!}{
        \includegraphics{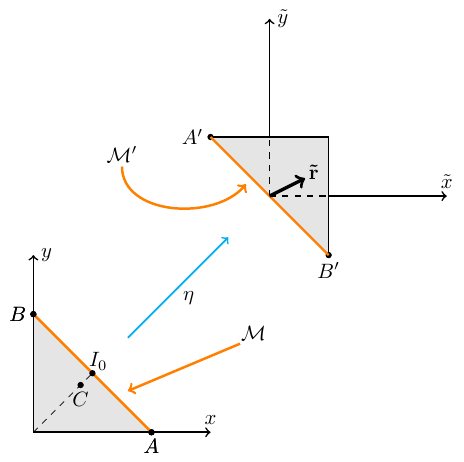}
    }
    \caption{
        A schematic view  of the diffeomorphism $\eta$    between the phase planes   $x$-$y$ and $\tx$-$\ty$.
        Bottom-left: 
            region in the $x$-$y$ phase plane accessible to the solutions of Eqs. \eqref{eq:autonomous1a},~\eqref{eq:autonomous1b}; 
            $A, B$, and $C$ are critical points ($C$ being present in the MNG only) and the subset $\mathcal{M}$ corresponding to $z=0$.
        Top-right: 
            corresponding region in the $\tx$-$\ty$ phase plane obtained through the diffeomorphism $\eta$, which also maps the critical points $A, B$ to $A', B'$ and the subset $\mathcal{M}$ to $\mathcal{M'}$.
        See text for details.
    }
    \label{fig:phaseSpace}
\end{figure}
% -------------------------------------------------------

\subsection{Estimate of the upper bounds}
\label{bounds}

Next, we provide for the two models approximate estimates of the maximum fractions of bilinguals that can emerge during the dynamics, $\zmax = \max\{ z\left(t\right): 0 <  t < t_{\rm conv} \}$, where $t_{\rm conv}$  denotes the convergence time, i.e., the time required for the system to reach consensus.

To this aim, we shall consider a coordinate transformation in the $x$-$y$ plane that transforms the ``old'' variables $x,y$ into ``new'' variables $\tx$, $\ty$:  $\left(\tx,\ty\right) = \eta\left(x,y \right) = (-x + 1/2,-y + 1/2)$. 
This is a linear diffeomorphism $\eta: \mathbb{R}^{2}\to \mathbb{R}^{2}$ that can be conveniently represented in matrix form,
\begin{equation}\label{eq:transform}
    \left(\begin{array}{c} \tx  \\ \ty  \end{array}\right)  = 
    \left(\begin{array}{cc} -1 & 0\\ 0 & -1 \end{array}\right)
    \left(\begin{array}{c} x \\ y \end{array}\right) 
    + \left(\begin{array}{c} 1/2 \\ 1/2 \end{array}\right)  \, ,
\end{equation}
equivalent to a  rotation plus a translation in the Cartesian plane.
It is also a linear isometric map -- indeed the transformation represented by the $2 \times 2 $ matrix in Eq.~\eqref{eq:transform} belongs to the special orthogonal group $SO(2)$ \cite{cornwell1997}.
Figure~\ref{fig:phaseSpace} illustrates how the transformation affects the original phase space,
%, i.e. the shadowed triangle in the lower part of the cartoon, in the Cartesian plane  $\left(x, y \right)$. 
showing also the position vector  $\mathbf{\tr}= (\tx,\ty)$:
the asymptotic stable equilibrium points $A = (1,0)$ and $B = (0,1)$ are mapped to $A' = (-1/2,1/2)$ and $B' = (1/2,-1/2)$, respectively, and the subset corresponding to  $z = 0$,
$\mathcal{M} = \{ (x,y) \IN \mathbb{R}^2: 0 \LE  x \leq 1,  0 \LE y \leq 1, x + y \EQ 1 \}$,
to
$\mathcal{M'} = \{ (\tx, \ty) \IN \mathbb{R}^2: |\tx| \LE 1/2,  |\ty| \LE 1/2, \tx \+ \ty \EQ 0 \}$.
Using the new variables $\tx, \ty$, and assuming that $p_A(t) \approx p_B(t) \equiv p(t)$, Eqs.~\eqref{eq:autonomous1a},~\eqref{eq:autonomous1b} become
\begin{align}
    \dot{\tx}     & = p \left(- \tx + 1/2 \right) \left(- \ty + 1/2 \right) - \left(\tx + \ty \right)^{2} 
              - \frac{ 3 - p}{2} \left(\tx + \ty \right) \left(- \tx + 1/2  \right) \, , 
    \label{eq:autonomous2a} 
\\[2ex]
    \dot{\ty}     & =  p \left(- \tx + 1/2 \right) \left(- \ty + 1/2 \right) - \left(\tx + \ty \right)^{2} 
              - \frac{ 3 - p}{2} \left(\tx + \ty \right) \left(- \ty + 1/2  \right) \, . 
    \label{eq:autonomous2b}
\end{align}
From these equations, noting that $z \equiv \tx + \ty$, one obtains 
\begin{equation}\label{eq:ode}
   \dot{z} = p \left(\frac{1}{2} + 2 \tx \ty - z \right) - 2 z^{2}  - \frac{ 3 - p}{2} z \left(1 - z \right) \, .
\end{equation}
This equation depends on the cross-term $ \tx \ty$, which can be rewritten as $\tx \ty = \left(z^{2} -  \tr^{2}\right)/2$, where  $\tr= \sqrt{\tx^{2}  + \ty^{2}}$ is the magnitude of the position vector $\mathbf{\tr}$ (see Fig.~\ref{fig:phaseSpace}).
Thus Eq.~\eqref{eq:ode} becomes
\begin{equation}\label{eq:ode1}
    \dot{z} = p \left( z^{2} - z + f \right) - 2 z^{2}  - \frac{ 3 - p}{2} z \left(1 - z \right) \, ,
\end{equation}
where we have defined the time-dependent function $f(t) = 1/2 - \tr^{2}(t)$.

Notice that Eq.~\eqref{eq:ode1} can be rewritten as 
\begin{equation}
    \label{eq:ode2}
    \dot{z} =  q_0( t ) + q_1( t ) z + q_2( t )  z^{2}   \, ,
\end{equation}
which is a first-order nonlinear differential equation of Riccati-type~\cite{ahmad2015, Jung2008} with coefficients  $q_0, q_1, q_2$, listed in Table \ref{table:table1} for the MNG and BNG.  
However, Eq.~\eqref{eq:ode2} is not strictly a Riccati equation, since the coefficient $q_0(t)$ depends on $z$.
%
% ----------------------------------------------
\begin{table}
\renewcommand{\arraystretch}{2} 
\caption{Coefficients  $q_0, q_1, q_2$  of the Riccati-type equation \eqref{eq:ode2} for the BNG and MNG models 
    (note that $q_0 \ge 0$, $q_1 <  0$,  and $q_2 \le 0$ for both models).
}
\label{table:table1}
%\begin{ruledtabular}
\begin{tabular}{c | c c c   }
Model & $q_0$ & $q_1$ &  $q_2$  \\ [1ex]
\hline
    BNG ($p_A=p_B=p$)   & $p f  $           & $-(3 + p)/2$          & $(p -1)/2$   \\
    MNG ($p=1$)         & $f$               & $- 2$                 & $ 0$    \\ 
%    2c-NG        & $f $          & $- (1 + 3 \beta)/2$        & $ (1 - \beta)/2$    \\ [1ex]
\end{tabular}
%\end{ruledtabular}
\end{table}
% ----------------------------------------------
%

The Cauchy problem defined by Eq.~\eqref{eq:ode2} with the initial condition $z_0 =0$, has a unique continuous solution $z(t)$, which  attains a global maximum at a certain time  $\Tt$, when $\dot{z}\left(\Tt\right) =0$, since  $z(t\to t_{\rm conv}) \to 0$.
In fact, for the case of the MNG model, i.e. when $p = 1$ in Eq.~\eqref{eq:ode2}, one obtains
\begin{equation}\label{eq:ode3}
 \dot{z} =   f - 2 z    \, .
\end{equation}
From here it is easy to find an upper bound $\zmax^\mathrm{MNG}$ for $z(t)$.
In the MNG, the system always reaches consensus and correspondingly $\mathbf{\tr}$ must reach either equilibrium point $A$ or $B$; therefore $ \tr^{2} \to 1/2$.
Thus the quantity $f$ is a non-negative bounded quantity, $0 \le f \le 1/2$. 
It follows that the maximum value of the solution occurs at some time $\Tt$  ($ 0< \Tt <t_{\rm conv} $) and satisfies the following inequality
\begin{equation}\label{eq:inequality}
    \zmax^\mathrm{MNG} = f/2 \lesssim 1/4  \, .
\end{equation}
This means that in the mean-field approximation the maximum number of bilingual agents that can emerge is always less than  $25\%$ of the overall population. 
We expect that this inequality overestimates the upper bound of the solution, as it usually happens in estimating \emph{a priori} the upper bound of the solution of an ordinary differential equation by means of the  Gr{\"o}nwall's lemma~\cite{Gronwall-1919a}. 
Indeed, in the MNG the presence of the critical point $C$ suggests that $\zmax^\mathrm{MNG}  \approx 0.236$, which is confirmed by our numerical results in Sec.~\ref{sec:results}.

For the case of the BNG model, the maximum of the solution $\zmax^\mathrm{BNG}$ is likely to be attained only in the phase space's region where  both  $\tx, \ty $  are non-negative and therefore the inequality $\tr^{2} \lesssim z^{2} $ holds.
Then, it is possible to show that the corresponding maximum fraction of bilinguals in the BNG is
\begin{equation}
\label{eq:inequality1}
    \zmax^\mathrm{BNG} \lesssim \frac{\tp}{\tp + 3}  \, ,
\end{equation}
where $\tp \equiv p\left( \Tt\right)$ denotes the generalization probability  at time $\Tt$.

The actual value of $\tp$ will depend on the chosen parameters of the Bayesian model. 
In the present case, the threshold probability value $p^*$ for generalizing the concept affects the value of $\tp$ and hence $\zmax^\mathrm{BNG}$.
Notice that since $p^* \leq \tp < 1$~\cite{Marchetti-2020a}, it follows that the upper limits defined in Eqs. \eqref{eq:inequality},~\eqref{eq:inequality1} fulfill the inequality
\begin{equation}
    \label{eq:inequality2}
    \zmax^\mathrm{BNG} < \zmax^\mathrm{MNG}  \, ,
\end{equation}
i.e., the maximum number of bilingual agents in the BNG model is always lower with respect to that observed in the MNG dynamics. 
Consistently, if $\tp \rightarrow 1$ (the learning process becomes a one-shot learning), then $\zmax^\mathrm{BNG} \to \zmax^\mathrm{MNG} \lesssim 1/4$.
The results about the maximum bilingual fractions are summarized in Table~\ref{table:table2}.

%
% ----------------------------------------------
\begin{table}
\renewcommand{\arraystretch}{2} 
\caption{Comparison of the upper limits of the bilinguals fractions for the MNG and BNG models.
The numerical result for the MNG model is estimated by integrating Eqs.~\eqref{eq:autonomous1a},~\eqref{eq:autonomous1b}. 
The results from the multi-agent simulations were obtained by averaging over $600$ realizations with $N=10^{4}$ agents; for the BNG the parameter values $\alpha=2$ and $p^* = 0.5$ were used. 
Note that according to Eq.~\eqref{eq:inequality1}, when  $\zmax^\mathrm{BNG}  \approx 0.20 $ one might expect $\tp \approx 0.7$. }
\label{table:table2}
%    \begin{ruledtabular}
    \begin{tabular}{c | c c c   }
    Model & Analytical results & Numerical result  & Simulations \\ [1ex]
    \hline
    MNG     &  $\zmax^\mathrm{MNG}  \lesssim 1/4$         & $\zmax^\mathrm{MNG}  \approx 0.236 $        & $\zmax^\mathrm{MNG}  \approx 0.236 $     \\

    BNG    &     $\zmax^\mathrm{BNG}  \lesssim \frac{\tp}{\tp + 3}$    & $-$         & $\zmax^\mathrm{BNG}  \approx 0.20 $      \\ [1ex]
    \end{tabular}
%\end{ruledtabular}
\end{table}
% ----------------------------------------------
%

% = = = = = = = = = = = = = = = = = = = = = = = = = = = = = = = 

\section{Dynamics of the multi-agent models} 
\label{sec:results}

The theoretical results presented in the previous section were obtained in the mean-field approximation, i.e. neglecting stochastic fluctuations, which are always present in multi-agent simulations~\cite{Stauffer-2007a,Vazquez-2010a,Baronchelli-2008a, Castellano-2009a} and play an important role as they induce a symmetry breaking of the Cauchy problem with symmetrical initial condition $I_0$.

% - - - - - - - - - - - - - - - - - - - - - - - - - - - - - - - - - - - - - - - - - - 

\subsection{Time evolution of the system}
\label{evolution}

In order to understand how the different learning processes in the MNG and BNG models affect the consensus dynamics, we study the time evolution of the fractions $x(t)$, $y(t)$, $z(t)$ and of the magnetization modulus $|m(t)|$ obtained from the simulation of the many-agent models.
Furthermore, we shall numerically validate the theoretical estimates of the maximum number of bilingual fractions. 
The results presented were obtained from simulations of systems of $N=5000$ agents, starting with the symmetrical initial condition $I_0=(x_0,y_0)=(0.5,0.5)$ (and $z(0)=0$) and averaging over $600$ dynamical realizations, unless indicated differently.
Introducing the magnetization $m(t) = x(t) - y(t)$, which measures the asymmetry between the monolingual communities using word A and B~\cite{Castello-2009a}, the initial condition $I_0$ corresponds to an unpolarized state characterized by zero magnetization and absence of bilinguals, i.e. $m(0)=0$, $z(0)=0$.

As discussed in Sec.~\ref{sec:meanfield}, in the MNG as well as in the BNG model a representative point starting from $I_0$ will remain on the line $x \!=\! y$ in the absence of fluctuations.
Taking an average of the population fractions over many independent runs would reproduce the mean-field trajectory, i.e. the trajectory with $x(t) \approx y(t)$ and magnetization $m(t) \approx 0$ starting from $I_0$ and converging to point $C$.
However, this is qualitatively different from any stochastic trajectory obtained from a many-agent simulation, in which the presence of fluctuations makes the trajectory starting from $I_0$ leave the $x=y$ line and eventually converge to a stable equilibrium point, either $A=(1,0)$ or $B=(0,1)$.

In order to gain information on the the time-dependence of the surviving monolingual fraction $x(t)$ (or $y(t)$) and of the disappearing monolingual fraction $y(t)$ (or $x(t)$) during the convergence to consensus, we partitioned the simulation runs into two sets, one with the trajectories converging toward point A and the other with trajectories converging to point B, selecting only one of the two sets for carrying out the ensemble averages.
Choosing the set of trajectories reaching the equilibrium point A, $x(t)$ represents the general time evolution of the surviving population fraction, while $y(t)$ that of the disappearing population fraction, so that one has a magnetization $m(t)>0$.
Notice that, in doing such an average, the average bilinguals fraction $z(t)$ remains unchanged, suggesting its general role within the MNG and BNG.

% -----------------------------------------------------------
\begin{figure}[ht!]
\resizebox{0.75\textwidth}{!}{
  \includegraphics{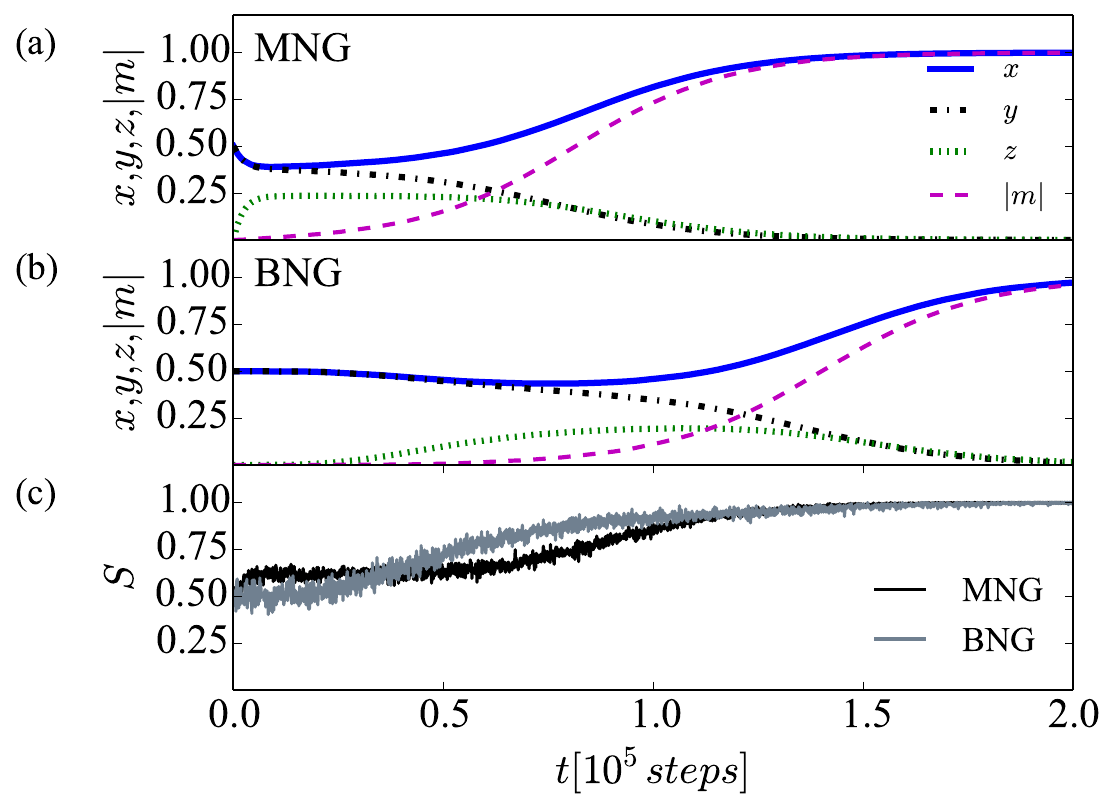}
}
    \caption{
      Time evolution of the fractions of agents $x, y$, $z$, and corresponding magnetization modulus $|m|=|x-y|$ for the MNG model (panel (a))   and BNG model with Erlang prior ($\alpha=2$) and $ p^*= 0.5$ (panel (b)).
      The curves were obtained from the realizations relaxing to the consensus corresponding to the equilibrium point $A=(x=1, y=0)$.
      Panel (c): Success rates $S(t)$ for the MNG and BNG model.
    }
\label{fig:curvature}
\end{figure}
% --------------------------------------------

In Fig.~\ref{fig:curvature} we plot the results for the MNG model (panel (a)) and BNG model with Erlang prior ($\alpha=2)$ and $ p^*= 0.5$ (panel (b)), obtained from runs that relax to the equilibrium point A, so that $x(t) \to 1$, $y(t) \to 0$, and $m(t) \to 1$ asymptotically.
In the beginning of the time evolution $x(t) \approx y(t)$ in  both models, but with an important difference:
in the MNG model, Fig.~\ref{fig:curvature}-(a), there is an initial fast decrease of $x(t), y(t)$ and a corresponding sudden increase of $z(t)$;
instead, Fig.~\ref{fig:curvature}-(b) shows that in the BNG model the system remains close to the initial condition $I_0$ for a much longer time-interval.
%(about $3 \times 10^4$ time-steps).

Thereafter, the curve $z(t)$ of the MNG model presents a plateau at $z \approx \zmax^\mathrm{MNG} \approx 0.23$.
Instead, the curve $z(t)$ of the BNG model exhibits a characteristic bell-shape with a more gradual increase and a maximum $\zmax^\mathrm{BNG} \approx 0.2$ reached at a larger time than in the MNG model. 
Notice that in both models the bilingual fraction $z$ is limited, as discussed in Sec.~\ref{sec:consideration}.
Furthermore, the values of the maxima $\zmax^\mathrm{MNG}$ and $\zmax^\mathrm{BNG}$ found from multi-agent simulations confirm the inequality \eqref{eq:inequality2} for systems of relatively large sizes.

The differences observed in the shapes of the curves at small and intermediate times are a direct consequence of the different underlying learning processes and can be understood by looking at the microscopic pairwise interactions between agents.

The initial fast increase of the bilinguals in the MNG model is due to the one-shot learning, in which a new name is immediately added to the hearer's list in a pairwise interaction as soon as the unknown word is conveyed for the first time.  
Instead, the multiple-iterate learning process of the BNG requires longer times for monolingual agents to learn a new word and to become bilingual, due to the need of recording a large enough number of positive examples, which significantly slows down the increase of the bilingual population~\cite{Marchetti-2020a}.

There are other relevant differences that affect the dynamics of the convergence to consensus, closely connected to the Bayesian nature of the BNG.
At a mean-field level they can be traced back to the time-dependence of the probabilities $p_A(t), p_B(t)$ and the fact that when $x(t) \ne y(t)$ also $p_A(t) \ne p_B(t)$, which introduces an asymmetry that has no analogue in the MNG model, recovered from the BNG model for $p_A = p_B = 1$.
The motion of the representative point in the $x$-$y$ plane in the direction perpendicular to the line $x=y$, toward either point A or B, starts as soon as stochastic fluctuations have induced a symmetry breaking leading to $x(t) \ne y(t)$ and is of particular importance for the study of consensus.
It can be illustrated effectively in terms of the magnetization rate $\dot{m} = \dot{x} - \dot{y}$, measuring how fast $x(t)$ and $y(t)$ become different from each other.
For the MNG model, setting $p_A = p_B = 1$ in Eqs.~\eqref{eq:autonomous1a}, \eqref{eq:autonomous1b}, one obtains a magnetization rate
$\dot{m} = m z$, showing that $m(t)$ grows proportionally with $m$ and with the same sign.
For the BNG model, the magnetization rate obtained from Eqs.~\eqref{eq:autonomous1a}, \eqref{eq:autonomous1b} for general $p_A(t) \ne p_B(t)$ is
$\dot{m} = \frac{1}{2}(3 - \bar{p}) m z - \frac{1}{2} \Delta p (1 - m^2 - z)$, where 
$\bar{p}(t) = \frac{1}{2}(p_A(t)+p_B(t))$ and 
$\Delta{p}(t) = p_B(t)-p_A(t)$.
The first term can be considered as the Bayesian counterpart of the MNG rate, with the crucial difference that it is time-dependent, starting from $\dot{m}(0)=0$ and only asymptotically tending to the MNG rate, $\dot{m}(t) \to m z$, as in that limit $\bar{p}(t) \to 1$.
The second term, proportional to the difference $\Delta p$, does not have a counterpart in the MNG model at all at any time $t$.
Of course, besides the motion perpendicular to the line $x=y$, the presence of different time-dependent probabilities also affect the motion parallel to the $x=y$ line, so that the resulting motion of the representative point in the BNG model presents the observed qualitative differences with respect to the MNG model.

Instead, as one can notice from Fig.~\ref{fig:curvature}, the dynamics of the convergence towards consensus are similar in the MNG and BNG models.
This can be understood noting that in this stage each Bayesian agent has already accumulated a relatively large number of examples for the unknown word. As a result, the generalization function $g \to 1$ and therefore also its ensemble averages $p_A,p_B \to 1$ in the evolution equations~\eqref{eq:autonomous1a},~\eqref{eq:autonomous1b}, which become equivalent to the mean-field equations of the MNG model.
The other process besides the word learning contributing to the dynamics is the agreement, which works in the same way in the BNG and MNG nodels. 
Thus, the dynamics of the two models become asymptotically equivalent.

To summarize, the curves of the the population fractions $x,y,z,m$ of the MNG and BNG differ first of all in the time scale of the convergence to consensus, as it is evident from Figs.~\ref{fig:curvature}-(a) and (b).
However, they also differ in some additional qualitative features of their shapes, which would persist even after a suitable scaling of the time variable.
Such major differences affect most of the time evolution of the system, becoming less pronounced only in the asymptotic stage of the relaxation to consensus.
The asymptotic behaviors of the two models present a similarity due to general constraint imposed by the set of trajectories selected, which reach consensus at point A: $x(t), m(t) \to 1$ while $y(t), z(t) \to 0$ for $t \to \infty$. 
Analogous asymptotic behaviors are shared by other three-state models.

The differences between the curves of the MNG and BNG discussed above are expected to have crucial consequences on the consensus dynamics of models describing the learning processes of multiple words characterized by different time scales (e.g. due to different levels of difficulty in learning the words) or in generalized models in which the weight of the role of bilinguals can be tuned through some parameters, see e.g. Ref.~\cite{Heinsalu-2014a}.

In order to monitor the evolution toward consensus, one can use the success rate $S(t)$, scoring one or zero at each time-step for a success or failure, respectively.
Figure~\ref{fig:curvature}-(c) shows the time evolutions of the success rates $S(t)$ for the MNG and BNG models. 
In the case of the MNG dynamics, the success rate $S$ exhibits a plateau analogously to the bilingual fraction $z$ in panel (a). 
These plateaus represent a reorganization phase of the system, in which the average rates of agreements and learning events are approximately constant.
Instead, in the BNG model, the small plateau observed at the beginning of the time evolution corresponds to a phase where  the majority of interactions consist of reinforcement processes.
After these initial phases, in both models the success rate $S$ grows monotonically.
In the later stage, the dynamics of convergence towards consensus become equivalent, as explained above.

% --------------------------------------------------------------------------- 
\begin{figure}[ht!]
\resizebox{0.75\textwidth}{!}{%
  \includegraphics{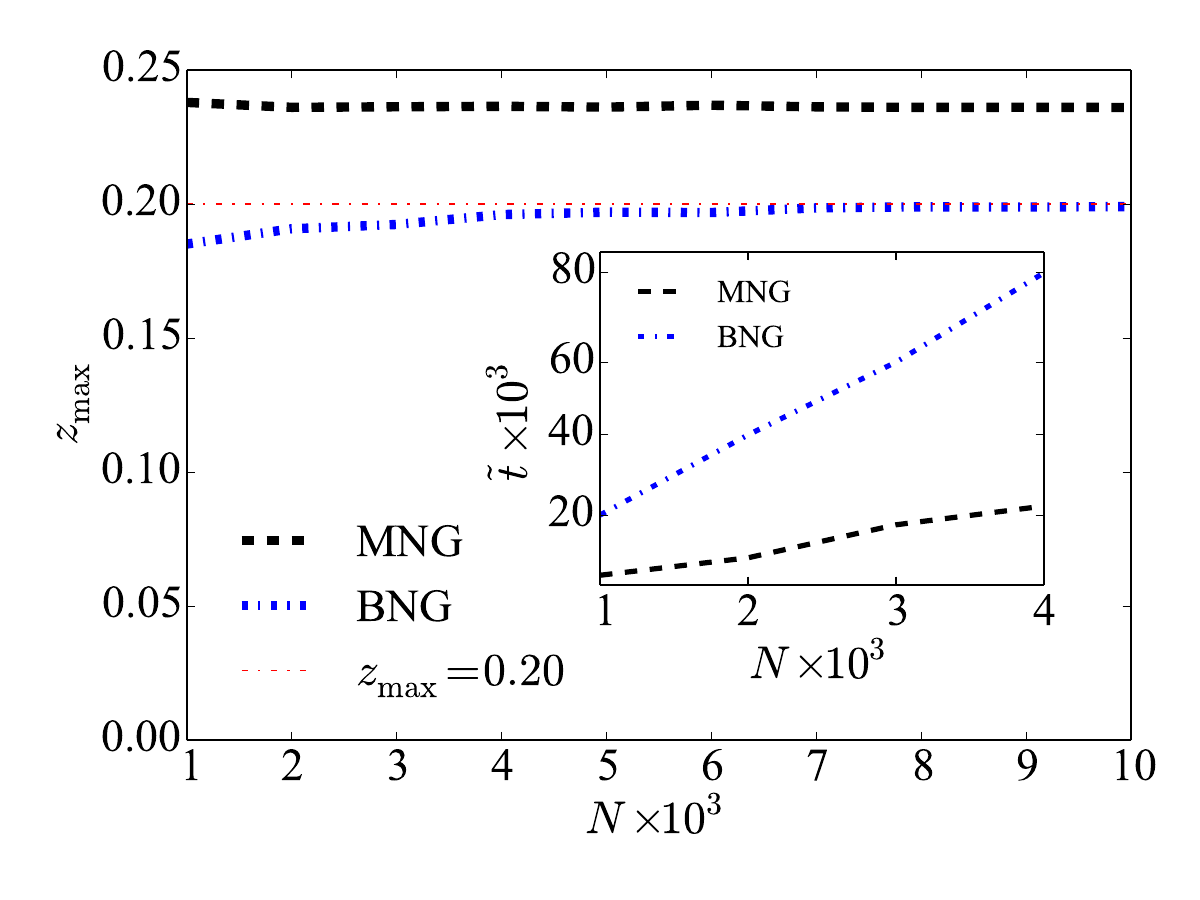}
}
\caption{
Maximum fraction of bilingual $\zmax$ as a function of the system size (or equivalently the number of agents) $N$ for the MNG model and the BNG model with the Erlang prior ($\alpha=2$) and  $ p^*= 0.5$. 
The curves are obtained averaging over $600$ runs starting from $m_0=0$.
Here $\zmax^\mathrm{MNG}  \approx 0.236$ and  $\zmax^\mathrm{BNG}  \approx 0.197$ for large $N$ i.e. in the asymptotic limit. 
Inset: corresponding time $\Tt$, at which $z(t) = \zmax$, as function of $N$. 
}
\label{fig:maxima}       
\end{figure}
% --------------------------------------------------------------------------- 

% --------------------------------------------------------------------------- 
\begin{figure}[ht!]
\resizebox{0.75\textwidth}{!}{%
  \includegraphics{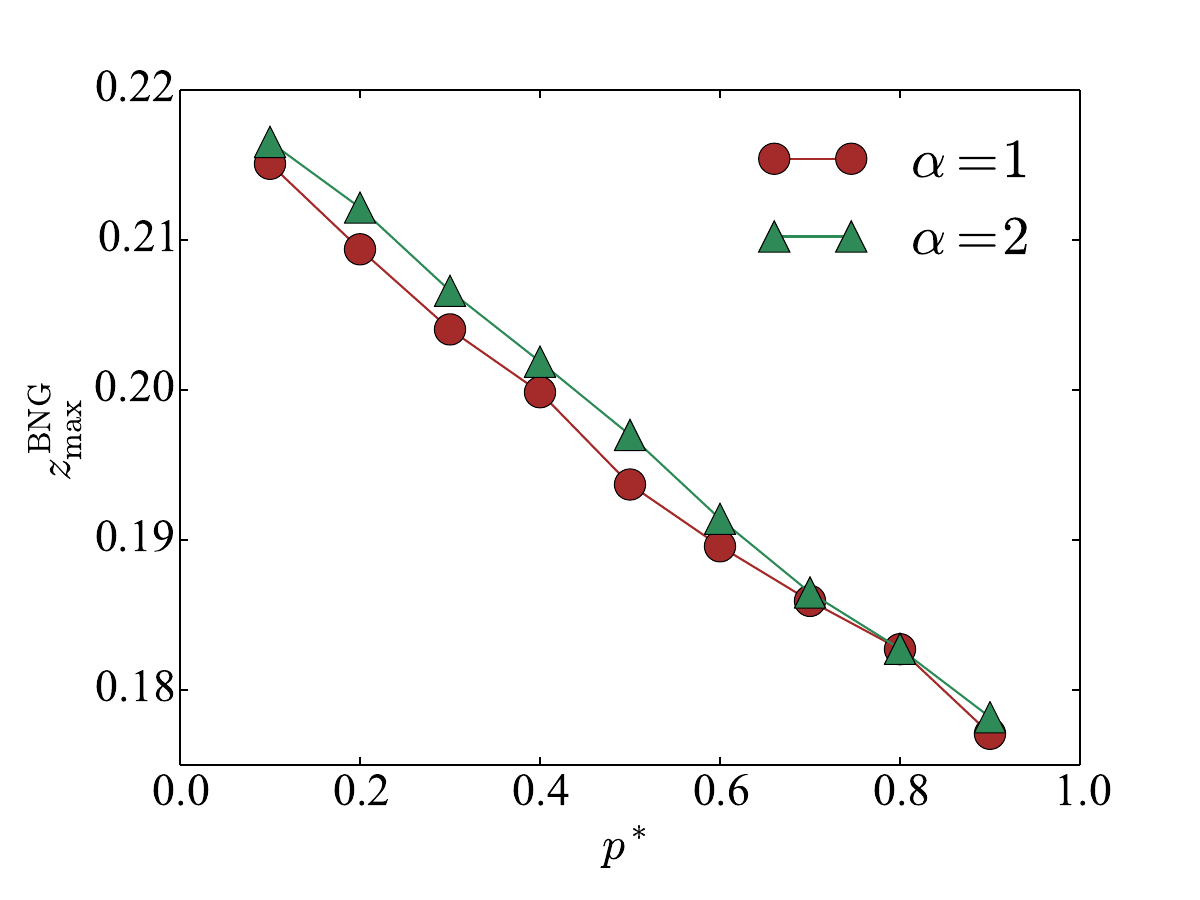}
}
\caption{
Comparison of the results for different priors and values of $p^*$: maximum number of bilingual $\zmax^\mathrm{BNG}$  as a function of the threshold generalization probability $p^*$ for the BNG with exponential prior (circles, $\alpha=1$) and Erlang prior (triangles, $\alpha=2$).
Lines are a guide to the eye.}
\label{fig:generalizationBoundary}      
\end{figure}
% --------------------------------------------------------------------------- 

% - - - - - - - - - - - - - - - - - - - - - - - - - - - - - - - - - - - - - - - - - - 

\subsection{Robustness of the results}
\label{sec:robust}

Next, we shall investigate the robustness of inequality~\eqref{eq:inequality2} when varying some parameters of the system.

We start by studying the dependence on the system size $N$. 
To this end, we carried out multi-agent simulations for various values of $N$ averaging over $600$ realizations.
In Fig.~\ref{fig:maxima} we present $\zmax$ as a function of system size $N$ in the range $N \in [10^3, 10^4]$. 
The values of $\zmax$ obtained are nearly independent of $N$ for both models, with $\zmax^\mathrm{MNG}  \approx 0.236$ for the MNG and $\zmax^\mathrm{BNG}  \approx 0.20$ for the BNG model; this trend is also confirmed by the simulations of systems with much larger size (not shown).
However, for the Bayesian model, a small deviation (1-7\%) from the asymptotic value $\zmax^\mathrm{BNG} \approx 0.20$ is observed when $N \lesssim 5000$. 
The observed discrepancy increases with decreasing the system size. The reason is that  some finite size effects strongly affect the time-dependence of  $p(t)$ ~\cite{Toral2007a, Brigatti-2016a}.
On the contrary,  no finite size effects are observed for the MNG model, not even for very small values of $N$. Note that despite these finite size effects, no violation of inequality~\eqref{eq:inequality2} is observed.

In the inset of  Fig.~\ref{fig:maxima}, we also plot the  time $\Tt$,  at which $z(t) = \zmax$, as a function of $N \in [10^3, 4 \times 10^3]$ for both models. 
We found that the curve corresponding to the BNG model exhibits a linear dependence on the system size, i.e. $\Tt \propto N$.

Thus, the mean-field inequality~\eqref{eq:inequality2} holds even in the presence of  stochastic fluctuations and time-dependent probabilities.
In fact, results concerning the bilingual fractions $z(t)$ should hold both in the many-agent model and in the mean-field limit because, as mentioned above, they are independent of the particular selection of realizations used for extracting mean values.

Let us turn our attention to the effects of the inductive biases in the BNG dynamics. 
This can be done either by tuning the probability threshold  $ p^*$ for a given prior or by choosing a different prior ($\alpha=1,2$). 
%First, let us assume that the agents, for some reasons, generalize at different values of $ p^*$ ($0<  p^* < 1$) for a given prior ($\alpha=1,2$). 
Figure~\ref{fig:generalizationBoundary} shows $\zmax^\mathrm{BNG}$ as function of $p^*$ for the Erlang prior ($\alpha=2$) and exponential prior ($\alpha=1$). 
The corresponding curves exhibit a monotonic behaviour, in which $\zmax^\mathrm{BNG}$ decreases in the range ($0.178, 0.216$), remaining below the value $\zmax^\mathrm{MNG}$, in agreement with inequality~\eqref{eq:inequality2}.
It is evident that for $p^* \rightarrow  0$ the Bayesian model becomes equivalent to the MNG model, as the characteristic one-shot learning is recovered. 
Since the two curves for different priors almost coincide, we can conclude that the choice of the prior is irrelevant for the problem at hand. 
This fact is not surprising because in the BNG model each Bayesian agent accumulates a large number of examples at convergence ($\bar{n}_{ex} \approx 80$).

% --------------------------------------------------------------------------- 
\begin{figure}[ht!]
\centering
\includegraphics[width=8.0cm]{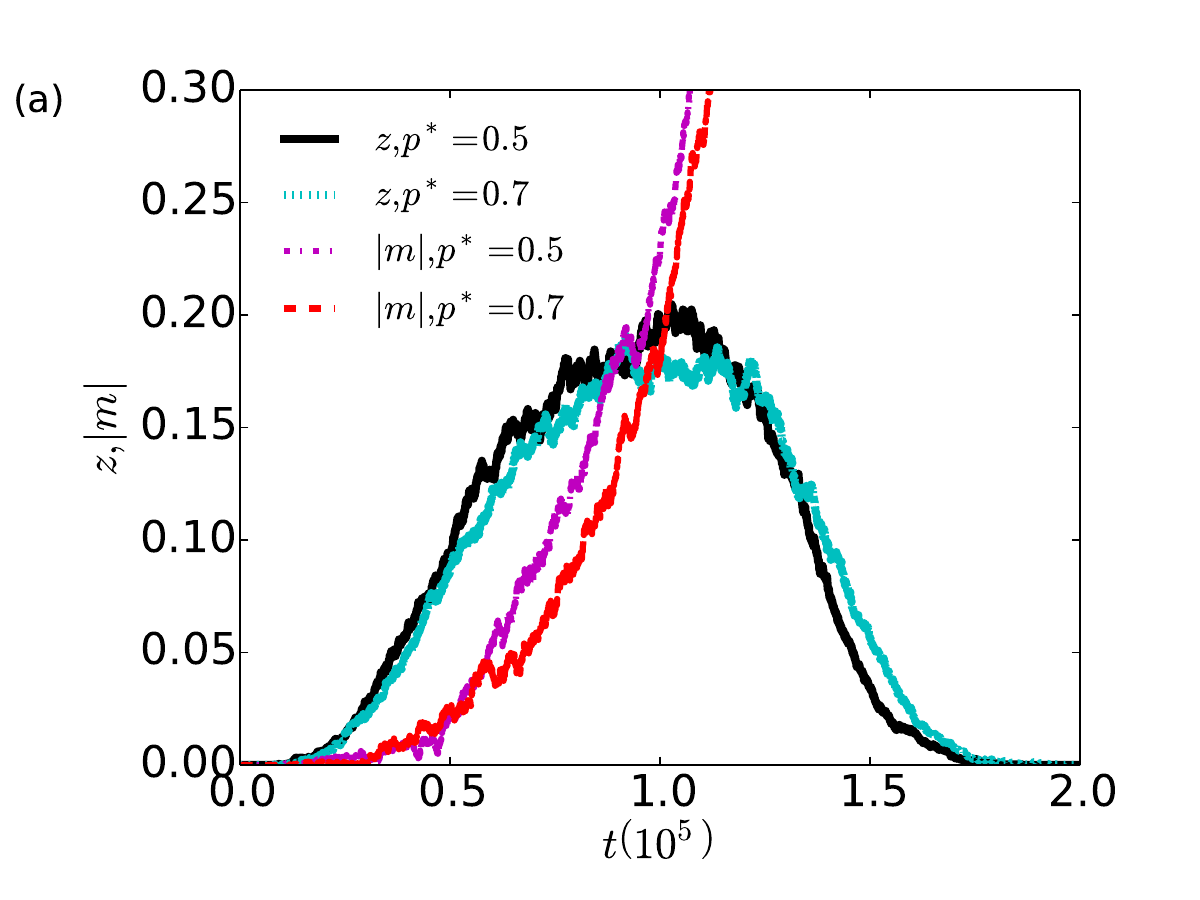}
\includegraphics[width=8.0cm]{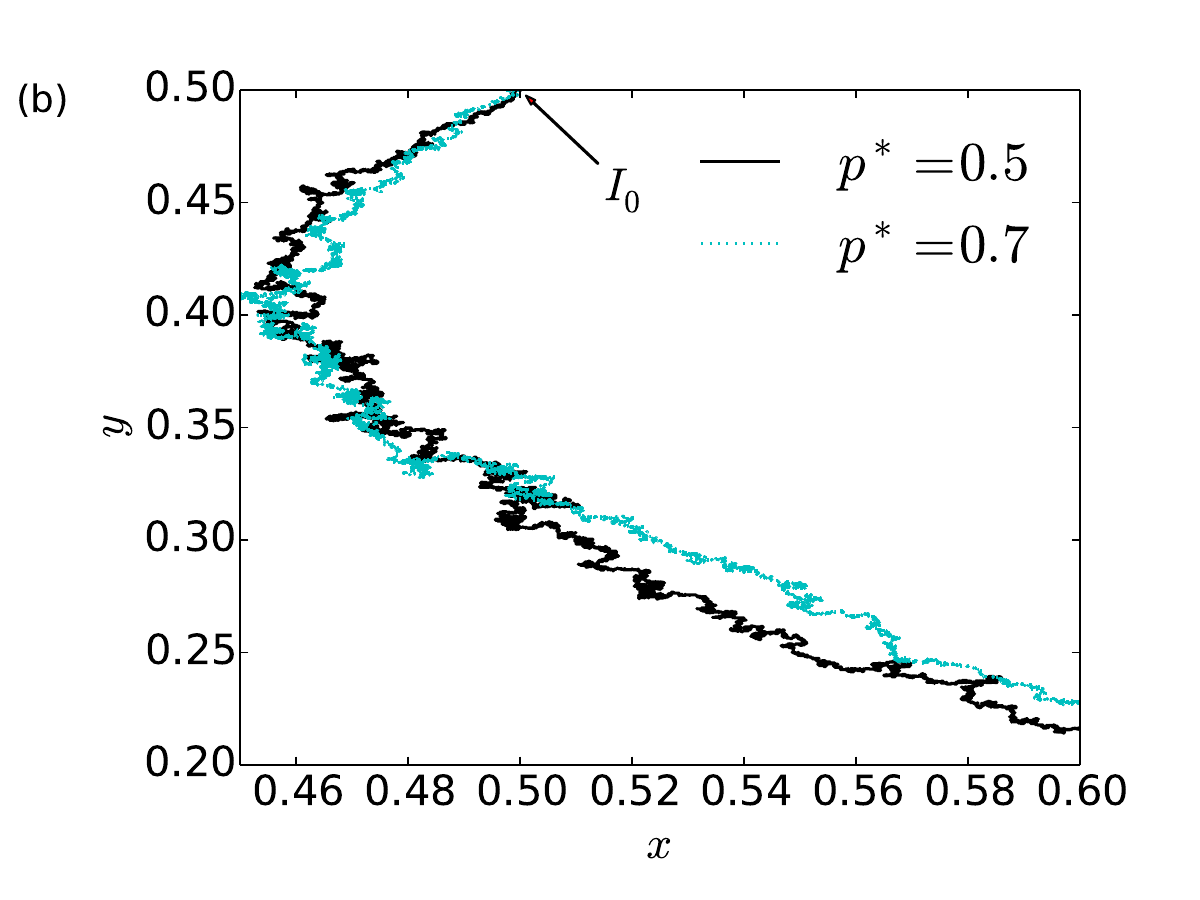}
\caption{
    Time evolution of the system from two runs of the BNG model with Erlang prior ($\alpha=2$), for $p^*=0.5$ and $p^*=0.7$.
    Panel (a):  Bilinguals fraction $z(t)$ and magnetization modulus $|m(t)|$ as a function of time. 
    Note that the curves $z(t)$ reach maximum values $\zmax^\mathrm{BNG} \approx 0.205$ and $\zmax^\mathrm{BNG} \approx 0.19$ for $p^*=0.5$ and $p^*= 0.7$, respectively.
    Panel (b): Early and intermediate stages of the corresponding trajectories in the $x$-$y$ plane emerging from the same initial condition $I_0$. 
}
\label{fig:bending}      
\end{figure}
% ----------------------------------------------------------------------------

Finally, Fig.~\ref{fig:bending} compares the results obtained from two runs of the BNG model using probability thresholds $p^*=0.5$ and $p^*= 0.7$.
Figure~\ref{fig:bending}-(a) depicts the time evolutions of the bilinguals fraction $z(t)$ and magnetization $m$;
the corresponding trajectories in the phase plane are shown in Fig.~\ref{fig:bending}-(b).
We expect that for the higher threshold, $p^*= 0.7$, generalization will be more difficult to achieve than for the lower threshold $p^* = 0.5$ and therefore more encounters will be needed in order to achieve consensus, resulting in a slower relaxation.
Even if the differences between the curves are small, nevertheless their time dependencies shown in Fig.~\ref{fig:bending}-(a) confirm a systematic trend to a slower convergence for the higher threshold $p^*= 0.7$; 
Notice that the maximum obtained for $p^*= 0.7$, $\zmax^\mathrm{BNG} \approx 0.190$, is smaller than that obtained for $p^*=0.5$, $\zmax^\mathrm{BNG} \approx 0.205$.
%Comparison of Fig.~\ref{fig:bending}-(a) and (b) suggests that the bending of the trajectory for $p^*=0.5$ starts before but remains smaller than that of the trajectory for $p^*=0.7$ in the first phase.
%This is consistent with the shape of the trajectories shown in Fig.~\ref{fig:bending}-(b), starting from the initial condition $I_0$ and following approximately the straight line $x=y$ ($m=0$) until  $t \approx 5\times 10^4$ steps: the trajectory corresponding to the larger threshold $p^*= 0.7$ starts to undergo a tighter bending, which can be associated to larger fluctuations in the magnetization $m$, as discussed above, preventing the trajectory from penetrating further into the permitted region of the phase plane (where $z$ is larger), which in turns leads to a smaller numerical value of $\zmax^\mathrm{BNG}$.

% = = = = = = = = = = = = = = = = = = = = = = = = = = = = = = = 

\section{Conclusion} 
\label{sec:conclusion}

In this paper we studied the BNG model, which is formally similar to the MNG model, but provides a more realistic picture of the word learning process by replacing the one-shot learning of the MNG with the learning process based on Tenebaum's Bayesian framework~\cite{Tenenbaum-1999,Perfors-2011a}.
In order to highlight the features of the BNG and clarify the role of the learning process on consensus dynamics, we carried out a systematic comparison of the results of the BNG model with those obtained from the MNG, which is a most simple model of semiotic dynamics, by performing numerical simulations of the multi-agent models and an analytical study of the corresponding mean-field equations.

We reformulated the mean-field equations of the two models in a unified way and, by introducing an approach based on the study of the fraction of bilinguals $z$, we could obtain analytical estimates of the upper limit $\zmax$ on the bilinguals fraction $z$, showing that $\zmax$ is always smaller in the BNG, with respect to the MNG.
Such a difference is one of the effects due to the presence of the learning process in the BNG.

In order to gain a better understanding of its origin and explore the other effects due to the underlying learning process, we carried out numerical simulations of the multi-agent models.
Both the MNG and its Bayesian counterpart asymptotically always converge towards a consensus state, a fact that makes the long-time stage of their time evolution similar.
The two models are characterized by different time scales of the relaxation to consensus, since the learning process in the BNG requires multiple interactions with other agents.
Furthermore, the dynamics at an early and intermediate stages are basically different as it is best seen by analyzing the time evolution of the fraction of bilinguals $z(t)$, for which the differences between the two models are more evident:
the MNG is characterized by a fast increase, followed by a plateau, in the number of bilinguals $z(t)$ versus time, due to the underlying one-shot learning; whereas in the BNG model $z(t)$ has a bell-shape with a well defined time scale $\tilde{t}$ at which $z(t)$ assumes its maximum value.

As a first step toward the understanding of how the qualitative differences between the BNG and the MNG are related to different underlying learning processes, we focused on the learning of a single word; however, such differences are expected to play a major role in more general models of consensus or opinion dynamics, e.g. with many words characterized by different learning time scales.
Furthermore, the difference between the time evolution of the bilingual fraction $z(t)$ can be relevant in language competition models where the influence of the bilingual population on the dynamics can be tuned through suitable parameters~\cite{Heinsalu-2014a}.

% = = = = = = = = = = = = = = = = = = = = = = = = = = = = = = = 

\section*{Acknowledgments}

The authors acknowledge support from the Estonian Ministry of Education and Research through Institutional Research Funding IUT39-1, 
the Estonian Research Council through Grant PUT1356 and PRG1059, 
and the ERDF (European Development Research Fund) CoE (Center of Excellence) program through Grant TK133. 

GM gratefully acknowledges useful comments from Daniel Siemssen and thanks Soon-Mo Jung and Themistocles Rassias for providing their paper on the Riccati equation.

% = = = = = = = = = = = = = = = = = = = = = = = = = = = = = = = 
%\section*{References}

\bibliography{references}

\begin{thebibliography}{10}
\expandafter\ifx\csname url\endcsname\relax
  \def\url#1{\texttt{#1}}\fi
\expandafter\ifx\csname urlprefix\endcsname\relax\def\urlprefix{URL }\fi
\expandafter\ifx\csname href\endcsname\relax
  \def\href#1#2{#2} \def\path#1{#1}\fi

\bibitem{Erlich2005}
P.~R. Ehrlich, S.~A. Levin, The evolution of norms, PLoS Biol 3 (2005) e--194.
\newblock \href {https://doi.org/10.1371/journal.pbio.0030194}
  {\path{doi:10.1371/journal.pbio.0030194}}.

\bibitem{Nyborg2016}
K.~Nyborg, J.~M. Anderies, A.~Dannenberg, T.~Lindahl, C.~Schill,
  M.~Schl{\"u}ter, W.~N. Adger, K.~J. Arrow, S.~Barrett, S.~Carpenter, F.~S.
  Chapin, A.-S. Cr{\'e}pin, G.~Daily, P.~Ehrlich, C.~Folke, W.~Jager,
  N.~Kautsky, S.~A. Levin, O.~J. Madsen, S.~Polasky, M.~Scheffer, B.~Walker,
  E.~U. Weber, J.~Wilen, A.~Xepapadeas, A.~de~Zeeuw, Social norms as solutions,
  Science 354~(6308) (2016) 42--43.
\newblock \href {https://doi.org/10.1126/science.aaf8317}
  {\path{doi:10.1126/science.aaf8317}}.

\bibitem{Baronchelli-2018a}
A.~Baronchelli, The emergence of consensus: a primer, R. Soc. open sci. 5
  (2018) 172189.
\newblock \href {https://doi.org/10.1098/rsos.172189}
  {\path{doi:10.1098/rsos.172189}}.

\bibitem{Baronchelli-2006c}
A.~Baronchelli, M.~Felici, V.~Loreto, E.~Caglioti, L.~Steels, Sharp transition
  towards shared vocabularies in multi-agent systems, J. Stat. Mech. (2006)
  P06014.

\bibitem{chen-2019a}
G.~Chen, Y.~Lou, Naming Game. Models, Simulations and Analysis, Springer
  International Publishing, Switzerland, 2019.

\bibitem{Wittgenstein-1953}
L.~Wittgenstein, Philosophical Investigations, Basil Blackwell, Oxford, UK,
  1986.

\bibitem{Steels-1995a}
L.~Steels, A self-organizing spatial vocabulary, Artif. Life 2 (1995) 319--332.

\bibitem{Steels-1997b}
L.~Steels, Language learning and language contact, in: W.~Daelemans, A.~Van~den
  Bosch, A.~Weijters (Eds.), Proceedings of the workshop on Empirical
  Approaches to Language Aquisition, 1997, pp. 11--24.

\bibitem{Baronchelli-2006a}
A.~Baronchelli, L.~Dall'Asta, A.~Barrat, V.~Loreto, Topology-induced coarsening
  in language games, Phys. Rev. E 73 (2006) 015102.
\newblock \href {https://doi.org/10.1103/PhysRevE.73.015102}
  {\path{doi:10.1103/PhysRevE.73.015102}}.

\bibitem{Baronchelli-2007a}
A.~Baronchelli, L.~Dall'Asta, A.~Barrat, V.~Loreto, The role of topology on the
  dynamics of the naming game, Eur. Phys. J. Spec. Top. 143 (2007) 233--235.
\newblock \href {https://doi.org/10.1140/epjst/e2007-00092-0}
  {\path{doi:10.1140/epjst/e2007-00092-0}}.

\bibitem{Baronchelli-2007b}
A.~Baronchelli, L.~Dall'Asta, A.~Barrat, V.~Loreto, Nonequilibrium phase
  transition in negotiation dynamics, Phys. Rev. E 76 (2007) 051102.
\newblock \href {https://doi.org/10.1103/PhysRevE.76.051102}
  {\path{doi:10.1103/PhysRevE.76.051102}}.

\bibitem{Marchetti-2020b}
G.~Marchetti, M.~Patriarca, E.~Heinsalu, A bird’s-eye view of naming game
  dynamics: From trait competition to {B}ayesian inference, Chaos: An
  Interdisciplinary Journal of Nonlinear Science 30 (2020) 063119.
\newblock \href {https://doi.org/10.1063/5.0009569}
  {\path{doi:10.1063/5.0009569}}.

\bibitem{Patriarca-2020a}
M.~Patriarca, E.~Heinsalu, J.~Leonard, Languages in Space and Time: Models and
  Methods from Complex Systems Theory, Physics of Society: Econophysics and
  Sociophysics, Cambridge University Press, 2020.

\bibitem{Baronchelli-2016a}
A.~Baronchelli, A gentle introduction to the minimal naming game, Belgian
  Journal of Linguistics 30~(1) (2016) 171--192.
\newblock \href {https://doi.org/10.1075/bjl.30.08bar}
  {\path{doi:10.1075/bjl.30.08bar}}.

\bibitem{Marchetti-2020a}
G.~Marchetti, M.~Patriarca, E.~Heinsalu, A {B}ayesian approach to the naming
  game model, Frontiers in Physics 8 (2020) 10.
\newblock \href {https://doi.org/10.3389/fphy.2020.00010}
  {\path{doi:10.3389/fphy.2020.00010}}.

\bibitem{Tenenbaum-1999}
J.~B. Tenenbaum, A {B}ayesian framework for concept learning, Ph.D. thesis, MIT
  (1999).

\bibitem{Griffiths-2007}
T.~L. Griffiths, M.~L. Kalish, Language evolution by iterated learning with
  bayesian agents, Cognitive Science 31~(3) (2007) 441--480.
\newblock \href {https://doi.org/10.1080/15326900701326576}
  {\path{doi:10.1080/15326900701326576}}.

\bibitem{Tenenbaum-1999b}
J.~B. Tenenbaum, {B}ayesian modeling of human concept learning, in: NIPS'98:
  Proceedings of the 11th International Conference on Neural Information
  Processing Systems, MIT Press, Cambridge, MA, USA, 1998, pp. 59--65.

\bibitem{Tenenbaum-2000b}
J.~B. Tenenbaum, F.~Xu, Word learning as {B}ayesian inference, in: Proceedings
  of the Annual meeting of the Cognitive Science Society, Vol.~22, 2000.

\bibitem{Tenenbaum2001}
J.~B. Tenenbaum, T.~L. Griffiths, Generalization, similarity, and {B}ayesian
  inference, Behavioral and brain sciences 24 (2001) 629--40; discussion
  652--791.
\newblock \href {https://doi.org/10.1017/S0140525X01000061}
  {\path{doi:10.1017/S0140525X01000061}}.

\bibitem{Xu-2007a}
F.~Xu, J.~B. Tenenbaum, Word learning as {B}ayesian inference, Psychological
  Review 114 (2007) 245--272.

\bibitem{Perfors-2011a}
A.~Perfors, J.~B. Tenenbaum, T.~L. Griffiths, F.~Xu, A tutorial introduction to
  {B}ayesian models of cognitive development., Cognition 120 (2011) 302--321.
\newblock \href {https://doi.org/10.1016/j.cognition.2010.11.015}
  {\path{doi:10.1016/j.cognition.2010.11.015}}.

\bibitem{Murphy-2012a}
K.~P. Murphy, Machine Learning: A Probabilistic Perspective, MIT Press,
  Cambridge, MA, 2012.

\bibitem{Tenenbaum-2011a}
J.~B. Tenenbaum, C.~Kemp, T.~L. Griffiths, N.~D. Goodman, How to grow a mind:
  Statistics, structure, and abstraction, Science 331 (2011) 1279--1285.
\newblock \href {https://doi.org/10.1126/science.1192788}
  {\path{doi:10.1126/science.1192788}}.

\bibitem{Lake2015}
B.~M. Lake, J.~B. Tenenbaum, R.~Salakhutdinov, Human-level concept learning
  through probabilistic program induction, Science 350~(6266) (2015)
  1332--1338.
\newblock \href {https://doi.org/10.1126/science.aab3050}
  {\path{doi:10.1126/science.aab3050}}.

\bibitem{Tversky1974}
A.~Tversky, D.~Kahneman, Judgment under uncertainty: Heuristics and biases,
  Science 185 (1974) 1124--1131.
\newblock \href {https://doi.org/10.1126/science.185.4157.1124}
  {\path{doi:10.1126/science.185.4157.1124}}.

\bibitem{Hahn2014}
U.~Hahn, The {B}ayesian boom: good thing or bad?, Frontiers in Psychology 5
  (2014) 765.
\newblock \href {https://doi.org/10.3389/fpsyg.2014.00765}
  {\path{doi:10.3389/fpsyg.2014.00765}}.

\bibitem{Sanborn2016}
A.~N. Sanborn, N.~Chater, {B}ayesian brains without probabilities, Trends in
  Cognitive Sciences 20~(12) (2016) 883--893.
\newblock \href {https://doi.org/https://doi.org/10.1016/j.tics.2016.10.003}
  {\path{doi:https://doi.org/10.1016/j.tics.2016.10.003}}.

\bibitem{Ngam2016}
V.~Ngampruetikorn, G.~J. Stephens, Bias, belief, and consensus: Collective
  opinion formation on fluctuating networks, Phys. Rev. E 94 (2016) 052312.
\newblock \href {https://doi.org/10.1103/PhysRevE.94.052312}
  {\path{doi:10.1103/PhysRevE.94.052312}}.

\bibitem{madsen2018}
J.~K. Madsen, R.~M. Bailey, T.~D. Pilditch, Large networks of rational agents
  form persistent echo chambers, Scientific Reports 8 (2018) 12391.
\newblock \href {https://doi.org/10.1038/s41598-018-25558-7}
  {\path{doi:10.1038/s41598-018-25558-7}}.

\bibitem{Jeffreys1961}
H.~Jeffreys, Theory of Probability, Clarendon Press, Oxford, 1939.

\bibitem{Baronchelli2010e}
A.~Baronchelli, T.~Gong, A.~Puglisi, V.~Loreto, Modeling the emergence of
  universality in color naming patterns, Proceedings of the National Academy of
  Sciences 107~(6) (2010) 2403--2407.
\newblock \href {https://doi.org/10.1073/pnas.0908533107}
  {\path{doi:10.1073/pnas.0908533107}}.

\bibitem{Loreto2012}
V.~Loreto, A.~Mukherjee, F.~Tria, On the origin of the hierarchy of color
  names, Proceedings of the National Academy of Sciences 109~(18) (2012)
  6819--6824.
\newblock \href {https://doi.org/10.1073/pnas.1113347109}
  {\path{doi:10.1073/pnas.1113347109}}.

\bibitem{Baronchelli-2015}
A.~Baronchelli, V.~Loreto, A.~Puglisi, Individual biases, cultural evolution,
  and the statistical nature of language universals: The case of colour naming
  systems, PLOS ONE 10~(5) (2015) 1--19.
\newblock \href {https://doi.org/10.1371/journal.pone.0125019}
  {\path{doi:10.1371/journal.pone.0125019}}.

\bibitem{Patriarca-2012a}
M.~Patriarca, X.~Castell{\'o}, J.~Uriarte, V.~Egu{\'i}luz, M.~{San Miguel},
  Modeling two-language competition dynamics, Adv. Comp. Syst. 15~(3\&4) (2012)
  1250048.

\bibitem{Castello2006a}
X.~Castell{\'o}, V.~M. Egu{\'i}luz, M.~S. Miguel, Ordering dynamics with two
  non-excluding options: bilingualism in language competition, New J. Phys. 8
  (2006) 306.

\bibitem{Hurford-1989a}
J.~Hurford, Biological evolution of the saussurean sign as a component of the
  language-acquisition device, Lingua 77 (1989) 187--222.

\bibitem{Nowak-1999a}
M.~A. Nowak, J.~B. Plotkin, D.~C. Krakauer, The evolutionary language game,
  Journal of Theoretical Biology 200~(2) (1999) 147 -- 162.
\newblock \href {https://doi.org/10.1006/jtbi.1999.0981}
  {\path{doi:10.1006/jtbi.1999.0981}}.

\bibitem{Edmonds2002}
P.~Edmonds, G.~Hirst, Near-synonymy and lexical choice share on, Computational
  Linguistics 28~(2) (2002) 105--144.
\newblock \href {https://doi.org/10.1162/089120102760173625}
  {\path{doi:10.1162/089120102760173625}}.

\bibitem{Stauffer-2007a}
D.~Stauffer, X.~Castell\'o, V.~M. Egu\'iluz, M.~San~Miguel, Microscopic
  {A}brams--{S}trogatz model of language competition, Physica A 374 (2007)
  835--842.
\newblock \href {https://doi.org/10.1016/j.physa.2006.07.036}
  {\path{doi:10.1016/j.physa.2006.07.036}}.

\bibitem{Vazquez-2010a}
F.~Vazquez, X.~Castell{\'o}, M.~{San Miguel}, Agent based models of language
  competition: macroscopic descriptions and order--disorder transitions, J.
  Stat. Mech. (2010) P04007\href
  {https://doi.org/10.1088/1742-5468/2010/04/P04007}
  {\path{doi:10.1088/1742-5468/2010/04/P04007}}.

\bibitem{cornwell1997}
J.~F. Cornwell, Group Theory in Physics. An Introduction, Academic Press,
  London, UK, 1997.

\bibitem{ahmad2015}
S.~Ahmad, A.~Ambrosetti, {A Textbook on Ordinary Differential Equations}, 2nd
  Edition, Springer International Publishing, Switzerland, 2015.

\bibitem{Jung2008}
S.-M. Jung, {\relax Th}.~M. Rassias, Generalized {H}yers-{U}lam stability of
  {R}iccati differential equation, Mathematical Inequalities \& Applications
  11~(4) (2008).
\newblock \href {https://doi.org/dx.doi.org/10.7153/mia-11-67}
  {\path{doi:dx.doi.org/10.7153/mia-11-67}}.

\bibitem{Gronwall-1919a}
T.~H. Gr{\"o}nwall, Note on the derivatives with respect to a parameter of the
  solutions of a system of differential equations, Annals of Mathematics 20
  (1919) 292–296.
\newblock \href {https://doi.org/10.2307/1967124} {\path{doi:10.2307/1967124}}.

\bibitem{Baronchelli-2008a}
A.~Baronchelli, V.~Loreto, L.~Steels, In-depth analysis of the naming game
  dynamics: The homogeneous mixing case, Int. J. Mod. Phys. C 19~(5) (2008)
  785--812.
\newblock \href {https://doi.org/10.1142/S0129183108012522}
  {\path{doi:10.1142/S0129183108012522}}.

\bibitem{Castellano-2009a}
C.~Castellano, S.~Fortunato, V.~Loreto, Statistical physics of social dynamics,
  Rev. Mod. Phys. 81 (2009) 591.
\newblock \href {https://doi.org/10.1103/RevModPhys.81.591}
  {\path{doi:10.1103/RevModPhys.81.591}}.

\bibitem{Castello-2009a}
X.~Castello, A.~Baronchelli, V.~Loreto, Consensus and ordering in language
  dynamics, Eur. Phys. J. B 71~(4) (2009) 557--564.
\newblock \href {https://doi.org/10.1140/epjb/e2009-00284-2}
  {\path{doi:10.1140/epjb/e2009-00284-2}}.

\bibitem{Heinsalu-2014a}
E.~Heinsalu, M.~Patriarca, J.~L. L{\'e}onard, The role of bilinguals in
  language competition., Advances in Complex Systems 17~(1) (2014) 1450003.
\newblock \href {https://doi.org/10.1142/S0219525914500039}
  {\path{doi:10.1142/S0219525914500039}}.

\bibitem{Toral2007a}
R.~Toral, C.~Tessone, Finite size effects in the dynamics of opinion formation,
  Comm. Comp. Phys. 2 (2007) 177--195.

\bibitem{Brigatti-2016a}
E.~Brigatti, A.~Hernandez, Finite-size scaling analysis of a nonequilibrium
  phase transition in the naming game model, Phys. Rev. E 94~(5) (2016) 052308.
\newblock \href {https://doi.org/10.1103/PhysRevE.94.052308}
  {\path{doi:10.1103/PhysRevE.94.052308}}.

\end{thebibliography}

% = = = = = = = = = = = = = = = = = = = = = = = = = = = = = = = 

\end{document}